\documentclass[11pt]{article}

\usepackage{graphicx}
\usepackage{amsmath}
\usepackage[utf8]{inputenc}
\usepackage{dsfont}
\usepackage{amsmath}
\usepackage{mathtools}
\usepackage{amssymb}
\usepackage{amsthm}
\usepackage[margin=0.95in]{geometry}
\usepackage{setspace}
\usepackage{float}
\usepackage{bm} 
\usepackage{soul}
\usepackage[dvipsnames]{xcolor} 
\usepackage{tikz}
\usepackage{pgfplots}
\pgfplotsset{compat=1.18}

\newcommand\R{\mathbb{R}}
\newcommand\E{\mathbb{E}}
\newcommand\Normal{\mathcal{N}}
\newcommand\eps{\varepsilon}

\newtheorem{theorem}{Theorem}
\newtheorem{corollary}{Corollary}
\newtheorem{lemma}{Lemma}
\newtheorem{proposition}{Proposition}
\newtheorem{definition}{Definition}
\newtheorem{assumption}{Assumption}

\usepackage[round,comma]{natbib}
\usepackage[hidelinks]{hyperref}
\hypersetup{
    colorlinks=false,
    linkcolor=blue,
    filecolor=blue,      
    urlcolor=blue,
    citecolor=black,
}

\title{Imprecision Attenuates Updating\footnote{I am grateful to Xiaosheng Mu for his advice on this project. For helpful discussions I thank Roland Bénabou, Christopher Chambers, Martin Cripps, Maxime Cugnon de Sévricourt, Loren Fryxell, Benjamin Enke, Faruk Gul, Navin Kartik, Alessandro Lizzeri, Pietro Ortoleva, Wolfgang Pesendorfer, and Fedor Sandomirskiy. I gratefully acknowledge financial support from the William S. Dietrich II Economic Theory Center. All errors are my own.}}
\author{Martin Vaeth\footnote{Paris School of Economics. Email: \href{mailto:martin.vaeth@psemail.eu}{\textcolor{blue}{martin.vaeth@psemail.eu}}}
}

\begin{document}

\maketitle
\thispagestyle{empty}
\setcounter{page}{0}

\begin{abstract}
\noindent 
This paper studies how imprecision in noisy signals attenuates Bayesian posterior means toward the prior mean, a property underlying many comparative-statics results in information economics. We introduce the precision order, characterized by the attenuation effect across symmetric location experiments: noise density $\tilde g$ is more precise than $g$ if and only if the posterior mean under $\tilde g$ is closer to the signal than under $g$ for all signal realizations and all symmetric, log-concave priors. We apply the precision order to derive comparative statics for prior precision, the value of information, and voting.
\vspace{8pt}

\textbf{Keywords:} signal extraction, underreaction, information orders, comparative statics, cognitive imprecision

\textbf{JEL Classifications:} C11, D83, D84
\end{abstract}

\newpage

\section{Introduction}
In many contexts, agents' responses to fundamentals are attenuated: consumers underreact to non-salient taxes (\citealp*{chetty_salience_2009}), numerical estimates are biased toward reference values (\citealp*{tversky_judgment_1974}), and, beyond economics, perceptual judgments regress toward the mean of the stimulus distribution (\citealp*{hollingworth_central_1910}; \citealp*{stevens_regression_1966}).

Such attenuated behavior is often explained through imprecise information. In the standard account, agents base their choices on noisy signals of the fundamentals, where the noise reflects imperfect observation of the state. A more recent literature on cognitive imprecision locates the noise in cognition instead: even when agents observe the fundamentals, the cognitive process mapping them to actions is noisy (\citealp*{woodford_modeling_2020}; \citealp*{enke_cognitive_2023}; \citealp*{ilut_economic_2023}). Both versions of the explanation share a core mechanism: noise attenuates Bayesian updating, compressing behavior toward a default action.

The imprecise-information explanation is usually formalized with the normal-normal model. The state $x$ is drawn from a normal prior with mean $\mu$, and the agent observes a signal $s = x + \eps$ with independent normal noise $\eps$. The posterior mean lies between the signal and the prior mean, and is farther from the signal for less precise signals (larger noise variance).\footnote{If $x \sim \Normal(\mu, \sigma_x^2)$ and $\eps \sim \Normal(0,\sigma_\eps^2)$ independent of $x$, then $\E[x\,|\,s]= \lambda\, s+\left(1-\lambda\right)\mu$ with $\lambda=\sigma_x^2/(\sigma_x^2+\sigma_\eps^2)$. For any signal realization $s$, larger $\sigma_\eps^2$ moves the posterior mean farther from the signal.} We call the comparative static in signal precision \textit{imprecision attenuates updating}, or \textit{attenuation effect} for short. The formalization is usually completed by assuming actions track the posterior mean, under which the attenuation effect carries over to behavior. A key piece of evidence for cognitive imprecision rests on the attenuation effect: subjects reporting higher cognitive uncertainty exhibit more attenuated behavioral responses (\citealp*{enke_cognitive_2023}; \citealp*{enke_behavioral_2024}).

Despite the theoretical and empirical relevance of the attenuation effect, it was unknown whether it extends beyond the normal distribution. Normality can be justified in particular cases, through the central limit theorem or in rational inattention with a normal prior and quadratic loss, but these cases are limited. Whether the effect is special to the normal distribution or holds more broadly therefore bears on how far the empirical evidence can be interpreted as a consequence of imprecise information. We characterize exactly when the attenuation effect holds and which assumptions drive it.

We prove the attenuation effect holds in all additive-noise models (location experiments) with symmetric noise under a symmetric, log-concave prior: for any signal realization, the posterior mean moves toward the signal realization as the signal becomes more precise (Theorem \ref{theorem1}). The conditions are nonparametric, replacing the normal functional form with two shape restrictions, symmetry and log-concavity. To state the result we introduce the \textit{precision order}. Noise density $\tilde g$ is more precise than noise density $g$ if the likelihood ratio $\tilde g(x)/g(x)$ decreases as $x$ moves away from 0, placing the absolute noise term of the more precise signal closer to 0 in the sense of likelihood-ratio domination (see Figure~\ref{fig:precise}). Scaling up a log-concave noise term produces a less precise signal under this definition.

The precision order is not only sufficient for the attenuation effect, but also necessary, so the effect characterizes the order. If noise density $\tilde g$ is \textit{not} more precise than noise density $g$, then for some symmetric, log-concave prior $f$ and some signal realization, $\tilde g$ induces \textit{more} attenuation than $g$. Moreover, the attenuation effect does not hold in general if we drop symmetry or log-concavity of the prior. 

Our main result extends the attenuation effect to a nonparametric class of distributions and pins down which distributional assumptions drive the effect. An essential assumption is symmetry, which is already needed for the more basic property that the posterior mean lies between the prior mean and the signal. \citet{chambers_updating_2012} show that without symmetric prior and noise densities, the posterior mean can easily fall outside this interval, so agents overreact or update in the opposite direction. 

More generally, the normal-normal model is pervasive in economics and is often used for comparative statics in signal precision. We show how our main result generates such comparative statics without parametric assumptions through several applications. We establish comparative statics for the median posterior mean conditional on the state, with an application to voting.\footnote{The Online Appendix includes comparative statics of the \textit{average} posterior mean conditional on the state.} We also study how prior precision, rather than signal precision, shapes updating. A more precise prior increases attenuation and lowers the value of information, the opposite of a more precise signal.

\paragraph{Related Literature} This paper introduces a new information order and provides comparative statics for signal structures with additive independent noise, called location experiments.

Prior work has provided orderings on location experiments related to the value of information. \citet{boll_comparison_1955} shows one location experiment Blackwell-dominates another if and only if the latter can be derived by adding independent noise to the former. This condition is very restrictive, making location experiments comparable only in rare cases. In response, \citet{lehmann_comparing_1988} introduced the less restrictive Lehmann order to rank location experiments by their value in all \textit{monotone} decision problems. We show the precision order strengthens the Lehmann order and neither implies nor is implied by the Blackwell order.

Other comparative statics results have been established for location experiments with log-concave noise distributions, which satisfy the monotone likelihood-ratio property. \citet{milgrom_good_1981} shows the strict monotone likelihood-ratio property is equivalent to higher signals being ``good news,'' in the sense of yielding first-order stochastically dominant posteriors, for any prior. Conversely, \citet{dawid_posterior_1973} examines violations of the ``good news''-property under heavy-tailed noise distributions: if the noise is sufficiently heavy-tailed relative to the prior, extreme observations are ``rejected,'' yielding posteriors that revert to the prior. Together with our results, this highlights log-concave location-scale experiments are a well-behaved class of signal structures that retains key properties of the normal-normal model. 

Other work studies properties of the distribution of posterior means induced by signal structures, with economic applications. \citet{kartik_information_2021} show that, under the monotone likelihood-ratio property and likelihood-ratio-ordered priors, each of two agents expects a more informative experiment to bring the other agent's posterior mean closer to her own prior mean. \citet{ganuza_signal_2010} order signals by the dispersion of the induced distribution of posterior means, whereas we order location experiments by the attenuation of the posterior mean toward the prior mean at each signal realization.

\section{Setup}
We study Blackwell experiments (signal structures), where the signal equals the one-dimensional state plus independent noise, called \textit{location experiments}. A Blackwell experiment is a family $(G_x(\cdot))_{x\in\R}$, where $G_x(\cdot)$ is the cumulative distribution function (CDF) over signals $s\in\R$ conditional on state $x$. A Blackwell experiment where $G_x(s)=G(s-x)$ for all $s,x\in\R$ is a location experiment. We restrict to location experiments where $G$ admits a density $g$ and assume the agent's prior admits a density.

Throughout the paper, $f,\tilde f,g,\tilde g$ denote densities, with $f,\tilde f$ prior densities and $g,\tilde g$ noise densities where the distinction matters. The posterior mean upon observing signal $s$ is
\[m(s,f,g):=\frac{\int x f(x)g(s-x)dx}{\int f(x)g(s-x)dx}.\]
We maintain the following regularity assumption.
\begin{assumption}\label{A1}
    All considered densities are strictly positive, bounded, continuously differentiable, and have finite first absolute moment.
\end{assumption}
Assumption~\ref{A1} ensures the signal has full support and $m(s,f,g)$ is well-defined for every $s$. The assumption rules out bounded-support densities such as the uniform, which we nonetheless use in several proofs. Our results extend to them: any uniform density is a pointwise limit of densities satisfying Assumption~\ref{A1} such that the posterior mean converges (Lemma~\ref{lemapprox}).

We will use the following assumptions on densities.
\begin{definition}
    The density $f$ is \emph{symmetric around $\mu$} if for all $x>0$, $f(\mu+x)=f(\mu-x)$. The density $f$ is \emph{symmetric} if there exists $\mu\in\R$ such that $f$ is symmetric around $\mu$.
\end{definition}
\begin{definition}
  The density $f$ is \emph{quasi-concave} if for all $x,x'\in\R$ and $\alpha\in(0,1)$, $f(\alpha x+(1-\alpha)x')\geq\min\{f(x),f(x')\}.$  
\end{definition}
\begin{definition}
   The density $f$ is \emph{log-concave} if $\log f$ is concave.
\end{definition}
Because we assume densities are strictly positive, log-concavity implies quasi-concavity.

The following result characterizes when the posterior mean lies between the signal and the prior mean, the property our comparative statics refine. For any $f$ let $\mu_f:=\int x f(x)dx$ denote the induced mean. All proofs are in the appendix.
\begin{proposition}[\citealp*{chambers_updating_2012}]\label{propch}
If the prior density $f$ is symmetric and quasi-concave and the noise density $g$ is symmetric around 0 and quasi-concave, then
\begin{align}\label{eqch}
\begin{split}
    &\forall s\leq \mu_f\colon \quad s\leq m(s,f,g)\leq \mu_f,\\
    &\forall s\geq \mu_f\colon \quad s\geq m(s,f,g)\geq \mu_f.
\end{split}
\end{align}
If \eqref{eqch} holds for all symmetric around 0 and quasi-concave $g$, then $f$ is symmetric and quasi-concave.\\ If \eqref{eqch} holds for all symmetric and quasi-concave $f$, then $g$ is symmetric around 0 and quasi-concave.
\end{proposition}
\citet{chambers_updating_2012} establish the first two statements. The third statement is new and follows from the duality \eqref{eqduality}. Symmetry and quasi-concavity of both densities are therefore necessary and sufficient for the posterior mean to lie robustly between the signal $s$ and the prior mean $\mu_f$, robustly meaning for every symmetric, quasi-concave density in the other role.

\section{Main Result}\label{secmainresult}
A challenge to formalizing the attenuation effect is finding the right order on location experiments. We introduce the \textit{precision order} and show it is characterized by the attenuation effect. We identify location experiments with their noise densities $g$ and restrict attention to symmetric noise densities.
\begin{definition}[Precision Order] \label{defprecise}
    Let $g$ and $\tilde g$ be symmetric around $\mu$. Then $\tilde g$ is \emph{more precise} than $g$ if the likelihood-ratio $\tilde g(x)/g(x)$ is nonincreasing in $x$ for $x>\mu$.
\end{definition}
\begin{figure}[t]
    \centering
   \begin{tikzpicture}
\begin{axis}[
  width=7.5cm, height=5.5cm,
  domain=-5:5, samples=200,
  xmin=-4, xmax=4, ymin=0, ymax=0.42,
  axis lines=middle,
  axis line style={-},
  xtick={0}, xticklabels={$\mu$},
  ytick=\empty,
  hide obscured x ticks=false,
  x axis line style={shorten >=-8pt, shorten <=-8pt},
  clip=false,
  xlabel style={font=\large}, tick label style={font=\large},
]
\addplot[black, thick, samples=200] {1/sqrt(2*pi)*exp(-x^2/2)}
  node[pos=0.52, right, yshift=5pt] {$\widetilde{g}$};
\addplot[black, thick, dashed, samples=200] {1/(1.5*sqrt(2*pi))*exp(-x^2/(2*1.5^2))}
  node[pos=0.52, right, yshift=-14pt] {$g$};
\end{axis}
\end{tikzpicture}
     \caption{Two normal densities $g$ and $\tilde g$ where $\tilde g$ is more precise.}
    \label{fig:precise}
\end{figure}
By the symmetry of densities, the definition implies the likelihood ratio ${\tilde g(x)/g(x)}$ is nondecreasing in $x$ for $x<\mu$. For noise densities symmetric around $0$, the precision order says $|\tilde\eps|$ is smaller than $|\eps|$ in the likelihood-ratio order, where $\eps\sim g$ and $\tilde\eps\sim\tilde g$. The more precise location experiment has a noise term closer to 0 in the sense of likelihood-ratio domination. Figure \ref{fig:precise} gives an example of two noise distributions ranked by the precision order. The more precise distribution has a higher density around zero, which falls faster as we move away from zero. Below we show the precision order has the intuitive property that scaling up a log-concave noise term makes it less precise, and we relate the precision order to existing information orders.

Our main result characterizes the precision order through the attenuation effect. Recall $\mu_f$ denotes the induced mean for any density $f$.
\begin{theorem}\label{theorem1}
For noise densities $g$ and $\tilde g$, the following are equivalent:
\begin{enumerate}
    \item Both $g$ and $\tilde g$ are symmetric around $0$, and $\tilde g$ is more precise than $g$.
    \item For every symmetric and log-concave prior density $f$: 
\begin{align}\label{eqordering}
\begin{split}
    \forall s\leq \mu_f\colon \quad &s\leq m(s,f,\tilde g)\leq m(s,f,g)\\
    \forall s\geq \mu_f\colon \quad &s\geq m(s,f,\tilde g)\geq m(s,f,g).
\end{split}
\end{align}
\end{enumerate}
\end{theorem}
The theorem shows the precision order is both sufficient and necessary for the attenuation ordering \eqref{eqordering}. For sufficiency, a more precise $\tilde g$ places the posterior mean closer to the signal realization $s$ than $g$ does, for every $s$ and every symmetric, log-concave prior, with both posterior means on the same side of $s$ as the prior mean $\mu_f$. Necessity shows the precision order cannot be weakened: if \eqref{eqordering} holds for every symmetric, log-concave prior, $g$ and $\tilde g$ are symmetric around $0$ and $\tilde g$ is more precise than $g$.

If the noise densities are also quasi-concave, Proposition~\ref{propch} places both posterior means weakly between the signal and the prior mean, so \eqref{eqordering} becomes \[s\le m(s,f,\tilde g)\le m(s,f,g)\le \mu_f\] for $s\le\mu_f$, and symmetrically for $s\ge\mu_f$. The less precise $g$ thus yields a posterior mean closer to the prior mean, for any signal.

To illustrate Theorem~\ref{theorem1}, consider an agent observing an economic forecast. The more precise she judges the forecast, the closer her posterior mean lies to the forecast. Section~\ref{secapplications} develops applications.

The difficulty in proving the result is that a more precise noise density has two opposing effects on the posterior mean. Consider $s>\mu_f$. The posterior $p(x)\propto f(x)\,g(s-x)$, so a more precise $\tilde g$ yields a posterior $\tilde p(x)\propto f(x)\,\tilde g(s-x)$ downweighting $x$ far from $s$ on \emph{both} sides of $s$. On $(-\infty,s]$, $\tilde p$ likelihood-ratio dominates $p$, moving the posterior mean up toward $s$, the desired direction. On $[s,\infty)$ the domination reverses, pulling the posterior mean down. The two effects oppose each other, so the result does not follow from standard monotone comparative statics.

The proof uses symmetry of the noise density to recast the posterior mean in a form where monotonicity results do apply: integrating a monotone function against the absolute noise term, which the precision order ranks in the likelihood-ratio order. We sketch the argument here and give the full proof in the appendix.

For sufficiency, let $s\ge \mu_f$, the case $s\le\mu_f$ being analogous by symmetry. Changing variables to $\eps := s - x$, splitting the integral at 0, and using $g(\eps)=g(-\eps)$,
\begin{align}\label{eqmthm1maintext}
    m(s,f,g) &= \frac{\int_{-\infty}^\infty (s-\eps) f(s-\eps) g(\eps)\,d\eps}{\int_{-\infty}^\infty f(s-\eps) g(\eps)\,d\eps} = s - \frac{\int_0^\infty \eps\,(f(s-\eps) - f(s+\eps))\,g(\eps)\,d\eps}{\int_0^\infty (f(s-\eps) + f(s+\eps))\,g(\eps)\,d\eps} \nonumber \\
    &= s - \frac{\int_0^\infty h(\eps,s)\,(f(s-\eps) + f(s+\eps))\,g(\eps)\,d\eps}{\int_0^\infty (f(s-\eps) + f(s+\eps))\,g(\eps)\,d\eps} = s - \int_0^\infty h(\eps,s)\,\nu_g(\eps)\,d\eps
\end{align}
where
\begin{equation}\label{eqh}
    h(\eps,s) := \eps\,\frac{f(s-\eps) - f(s+\eps)}{f(s-\eps) + f(s+\eps)}
\end{equation}
and $\nu_g$ is the density on $[0,\infty)$ proportional to $(f(s-\eps)+f(s+\eps))\,g(\eps)$. The proof establishes $h(\eps,s)$ is nondecreasing in $\eps$ on $[0,\infty)$ using symmetry and log-concavity of $f$. If $\tilde g$ is more precise than $g$, then $g$ likelihood-ratio dominates $\tilde g$ on $[0,\infty)$, and multiplying both noise densities by $f(s-\eps)+f(s+\eps)$ preserves this. Hence $\nu_g$ likelihood-ratio dominates $\nu_{\tilde g}$, and in particular first-order stochastically dominates it. Integrating the nondecreasing $h$ then yields a larger integral under $\nu_g$ than under $\nu_{\tilde g}$, which by \eqref{eqmthm1maintext} gives $m(s,f,g)\le m(s,f,\tilde g)$.

For necessity, symmetry of $g$ and $\tilde g$ around $0$ follows as in the proof of Proposition~\ref{propch}. Given symmetry, suppose $\tilde g$ is not more precise than $g$, so $\tilde g(x)/g(x)$ is strictly increasing at some $x>0$. Under a uniform prior on a small interval around $0$ and the signal $s=x>0=\mu_f$, the posterior under $g$ likelihood-ratio dominates the posterior under $\tilde g$, so $m(s,f,\tilde g)<m(s,f,g)$, reversing \eqref{eqordering}. Both steps use only uniform priors, so \eqref{eqordering} holding for any prior class containing the uniform densities already forces $g$ and $\tilde g$ to be symmetric around $0$ with $\tilde g$ more precise than $g$.

The two assumptions on the prior, symmetry and log-concavity, enter the proof of Theorem~\ref{theorem1} only through properties of $h(\eps,s)$. For $s\ge\mu_f$, whenever $h(\cdot,s)$ is nonnegative and nondecreasing on the noise support, the argument delivers \eqref{eqordering} at that signal, and symmetrically for $s\le\mu_f$. These properties depend on the prior and the signal but not on the noise densities. The ordering \eqref{eqordering} can hold at some signals with neither symmetry nor log-concavity of the prior. However, Appendix~\ref{app:prior} shows symmetry is necessary if we require the ordering at every signal, and log-concavity cannot be weakened to quasi-concavity. 

Theorem~\ref{theorem1} shows the familiar ordering \eqref{eqordering} holds far more generally than for the normal distribution. No parametric assumption is needed, only common distributional ones. The effect is thus robust across functional forms, but symmetry cannot be relaxed, which is a strong assumption. An asymmetric prior or noise density may still yield \eqref{eqordering} at some $s$, especially when the asymmetry is mild, but not in general. The symmetry requirement is not specific to our comparative statics. Already the property that the posterior mean lies robustly between the prior mean and the signal realization requires symmetry, as Proposition~\ref{propch} shows.

\subsection{Properties of the Precision Order}\label{secproperties}

Scaling up noise with density $g$ by $\sigma$ yields the density $g_\sigma:=g(\cdot/\sigma)/\sigma$, which is less precise under a mild condition.
\begin{lemma}\label{lemscale}
Let $g$ be symmetric around $0$. Then $g_\sigma$ is less precise than $g$ for all $\sigma>1$ if and only if $g(e^x)$ is log-concave. In particular, $g(e^x)$ is log-concave if $g(x)$ is log-concave.
\end{lemma}
The condition in Lemma~\ref{lemscale} is easy to verify and holds broadly. All location experiments with symmetric, log-concave noise densities are covered, including the normal, logistic, Laplace, and uniform.\footnote{\label{footnoteuniform} The uniform density has bounded support and so violates Assumption~\ref{A1}. However, a uniform density can be approximated by a log-concave density with full support (Lemma~\ref{lemapprox}), so our result extends to scaling uniform noise.} The Online Appendix gives examples of non-log-concave densities $g$ for which $g(e^x)$ is log-concave.

Noise densities $g$ symmetric around $0$ with $\log g(e^x)$ concave are quasi-concave.\footnote{With $\varphi(x):=\log g(e^x)$ concave, $\varphi'$ is nonincreasing, and $\varphi'(x)=e^x g'(e^x)/g(e^x)\to 0$ as $x\to-\infty$ because $g(0)>0$ (Assumption~\ref{A1}). Hence $\varphi'\le 0$, so $g$ is nonincreasing on $(0,\infty)$ and thus quasi-concave.} Combining Theorem~\ref{theorem1} with Proposition~\ref{propch}, a larger scale parameter moves the posterior mean away from the signal toward the prior mean, for every such noise density. Because scaling normal noise is the standard way agents vary signal precision in information-acquisition problems (e.g., \citealp*{van_nieuwerburgh_information_2009}), Lemma~\ref{lemscale} opens a route to study such problems beyond normal distributions.

The precision order neither implies nor is implied by the Blackwell order. Blackwell domination is very restrictive for location experiments: by \citet{boll_comparison_1955}, one location experiment Blackwell-dominates another if and only if the latter's noise term is obtained from the former's by adding independent noise. By this characterization, a location experiment with noise term $\tilde\eps\sim U[-1,1]$ Blackwell-dominates one with noise term $\eps\sim U[-c,c]$ only when $c$ is an integer. The precision order instead ranks $U[-1,1]$ as more precise than $U[-c,c]$ for every $c>1$ (see footnote~\ref{footnoteuniform}), as it does any scaling of a symmetric log-concave density. Hence the precision order does not imply the Blackwell order.

Perhaps surprisingly, the Blackwell order does not imply the precision order. For a counterexample, let $\tilde\eps=\eps+\eps'$ be the independent sum of a Student-$t$ noise term $\eps$ with parameter $\nu=2$ and $\eps'\sim U[-1,1]$. The experiment with noise $\eps$ Blackwell-dominates the experiment with noise $\tilde\eps$, yet the density of $\tilde\eps$ is not less precise than the one of $\eps$. By Theorem~\ref{theorem1}, Blackwell domination is therefore not sufficient for the attenuation ordering \eqref{eqordering}.

The precision order strengthens the Lehmann order on log-concave location experiments, which ranks experiments by their value in all \emph{monotone} decision problems (\citealp*{lehmann_comparing_1988},\citealp*{quah_comparative_2009}). The Lehmann order is defined for location experiments only under log-concave noise, the case satisfying the monotone likelihood-ratio property. Such an experiment dominates another in the Lehmann order if its noise is smaller in the dispersive order, meaning any two quantiles are weakly closer: CDF $\tilde G$ is smaller than CDF $G$ in the dispersive order if
\begin{equation}\label{eqdispersive}
\forall\ 0<\alpha<\beta<1:\ \tilde G^{-1}(\beta)-\tilde G^{-1}(\alpha)\leq G^{-1}(\beta)-G^{-1}(\alpha),
\end{equation}
where $G^{-1}$ and $\tilde G^{-1}$ are the right-continuous inverses of $G$ and $\tilde G$. Inequality~\eqref{eqdispersive} holds whenever $\tilde g=\tilde G'$ is more precise than $g=G'$. Truncated to positive values the precision order gives the likelihood-ratio order, which implies the hazard-rate order (\citealp*{shaked_stochastic_2007}, Theorem 1.C.1), which under log-concavity implies the dispersive order (\citealp*{bagai_tail-ordering_1986}, Theorem 1). Symmetry extends the inequality from $1/2\leq\alpha<\beta<1$ to all $0<\alpha<\beta<1$.

\section{Applications}\label{secapplications}
We apply Theorem~\ref{theorem1} to comparative statics for the median posterior mean, voting, prior precision, and the value of information.

\subsection{Median Posterior Mean and Voting}\label{secmedian}
In many settings the researcher observes the state $x$ and the population distribution of posterior means, but not the individual signals. For example, the researcher observes realized inflation together with the forecasts made a period earlier. The median of such a population admits a clean characterization under symmetric log-concave noise. We treat the average posterior mean in the Online Appendix.

Let $\widehat m(x,f,g)$ denote the median posterior mean under state $x$, the median of $m(s,f,g)$ under the signal density $g(\cdot-x)$. Taking the agents' signals to be independent conditional on the state, the empirical median of their posterior means converges to $\widehat m(x,f,g)$ as the population grows large. When $g$ is symmetric around $0$, the median signal given state $x$ equals $x$. Under a strictly log-concave noise density $g$ the location experiment satisfies the strict monotone likelihood-ratio property, so $m(s,f,g)$ is strictly increasing in $s$ and $\widehat m(x,f,g)=m(x,f,g)$. This identity reduces the next result to Theorem~\ref{theorem1} at $s=x$.

\begin{corollary}\label{cormedian}
Let the noise densities $g$ and $\tilde g$ be symmetric around $0$ and strictly log-concave. Then $\tilde g$ is more precise than $g$ if and only if for every symmetric and log-concave prior density $f$:
\begin{align}\label{eqmedian}
\begin{split}
    \forall x\leq \mu_f\colon \quad &x\leq \widehat m(x,f,\tilde g)\leq \widehat m(x,f,g)\le\mu_f\\
    \forall x\ge \mu_f\colon \quad &x\ge \widehat m(x,f,\tilde g)\ge \widehat m(x,f,g)\ge\mu_f.
\end{split}
\end{align}
\end{corollary}
\begin{proof}
    The identity $\widehat m(x,f,g)=m(x,f,g)$ turns \eqref{eqmedian} into \eqref{eqordering} at every signal $s=x$, with the outer bounds from Proposition~\ref{propch}. Both directions then follow from Theorem~\ref{theorem1}.
\end{proof}

In the forecasting example, a more precise signal noise density brings the median forecast closer to realized inflation, for every symmetric log-concave prior density held by forecasters. The result requires no parametric assumptions and therefore provides a better theoretical foundation for interpreting attenuated responses as a consequence of imprecise information.

The median posterior mean is more robust to the mapping from beliefs to actions than the average posterior mean, another common statistic. When the action is an increasing, possibly nonlinear, function $\varphi$ of the posterior mean, the median action equals $\varphi(\widehat m(x,f,g))$, so Corollary~\ref{cormedian} carries over to the median action. An analogous result on the average posterior mean would carry over to the average action only for affine $\varphi$, since averaging commutes with $\varphi$ only then. Moreover, the median action remains well-defined under a purely ordinal action space.

Our median characterization speaks to common-interest elections under sincere voting. A large electorate chooses between two fixed alternatives $L<R$ in $\mathbb{R}$ by majority rule. Voters share the spatial preference $u(a,x)=-(a-x)^2$ over the elected alternative $a\in\{L,R\}$, given state $x\in\mathbb{R}$. Each voter observes a private signal $s_i=x+\varepsilon_i$ with symmetric, strictly log-concave noise density $g$ and holds the symmetric, log-concave prior density $f$. A sincere voter elects $a\in\{L,R\}$ to maximize the expectation of $u(a,x)$ given her posterior over $x$, without conditioning on the event that her vote is pivotal.\footnote{Under strategic voting, comparative statics in precision are uninteresting: an equilibrium exists in which information is aggregated efficiently and the alternative closer to $x$ wins with probability approaching one as the electorate grows, irrespective of precision (\citealp*{mclennan_consequences_1998}). \citet{esponda_hypothetical_2014} find a majority of subjects vote non-strategically in a common-interest voting experiment.} Thus, she votes for $R$ whenever $m(s_i,f,g)>c$ with $c:=(L+R)/2$. As the electorate grows, the majority elects $R$ if and only if $\widehat m(x,f,g)>c$, assuming ties are broken toward $L$. Attenuation of $\widehat m(x,f,g)$ toward $\mu_f$ biases the election toward the alternative closer to the prior mean, so the election fails to aggregate information efficiently. Suppose $\mu_f<c$ and let $x^\ast>c$ satisfy $\widehat m(x^\ast,f,g)=c$. On every state $x\in(c,x^\ast)$ the alternative closer to the state is $R$, yet the majority elects $L$ because $\widehat m(x,f,g)<c$, so elections over-select the alternative closer to the prior mean. By Corollary~\ref{cormedian} a more precise noise density shrinks $(c,x^\ast)$ and thereby reduces the set of states on which the electorate errs. Perceived precision has the same effect. Suppose the objective noise density $g$ is symmetric around $0$ and voters update as if their noise density were $\hat g$, symmetric and strictly log-concave. The median signal is then $x$ and the median posterior mean $m(x,f,\hat g)$. By Corollary~\ref{cormedian}, a more precise perceived $\hat g$ shrinks the error set $(c,x^\ast)$, so overconfident voters err on fewer states.

\subsection{Prior Precision and Value of Information}
Location experiments exhibit a useful duality between the prior and the noise density, captured by
\begin{equation}\label{eqduality}
m(s,f,g)=s-m(s,g,f).
\end{equation}
Applying Theorem~\ref{theorem1} with the roles of prior and noise interchanged yields a dual result for making the prior more precise.
\begin{corollary}\label{dualcorollary}
For any $f$, $\tilde f$, and $\mu\in\R$, the following are equivalent:
\begin{enumerate}
    \item Both $f$ and $\tilde f$ are symmetric around $\mu$, and $\tilde f$ is more precise than $f$.
    \item For every symmetric around 0 and log-concave $g$: 
\begin{align}\label{eqorderingprior}
\begin{split}
    \forall s\leq \mu\colon \quad &m(s,f,g)\leq m(s,\tilde f,g)\leq \mu\\
    \forall s\geq \mu\colon \quad & m(s,f,g)\geq m(s,\tilde f,g)\geq \mu.
\end{split}
\end{align}
\end{enumerate}
\end{corollary}
Corollary~\ref{dualcorollary} shows a more precise prior moves posterior means closer to the prior mean, the opposite direction to a more precise noise density. We now turn to a consequence for the value of information. Classic orders such as the Blackwell and Lehmann orders compare signal structures for a fixed prior. We instead fix the signal structure and vary the prior, and show in mean-measurable decision problems a more precise prior lowers the value of every location experiment with symmetric, log-concave noise.

The agent chooses an action $a\in\mathcal A\subseteq\R$ from a compact set, with utility
\[
  u(a,x)=\alpha(a)+\beta(a)\,x+\gamma(x),
\]
where $\alpha,\beta,\gamma\colon\R\to\R$ are continuous and $\int |\gamma(x)|f(x)dx<\infty$. The agent's prior over the state $x$ has density $f$ with mean $\mu$.

For example, estimating $x$ under quadratic loss $u(a,x)=-(a-x)^2$ corresponds to $\alpha(a)=-a^2$, $\beta(a)=2a$, $\gamma(x)=-x^2$. Deciding whether to buy a good of quality $x$ at price $p$ corresponds to $\mathcal A=\{0,1\}$ and $u(a,x)=(x-p)\,a$, that is $\alpha(a)=-p\,a$, $\beta(a)=a$, $\gamma(x)=0$.

The value of a location experiment with noise density $g$ under prior $f$ is the expected utility attainable with the location experiment minus the expected utility attainable without it,
\[
  v(f,g):=\max_{\hat a\colon\R\to\mathcal A} \int\!\!\int u(\hat a(s),x)\,g(s-x)f(x)\,ds\,dx -\max_{a\in\mathcal A}\int u(a,x)f(x)\,dx.
\]
Because the interaction term $\beta(a)\,x$ in $u(a,x)$ is linear in the state, $v(f,g)$ depends on the experiment only through the induced distribution of posterior means, as is well known (and derived in the appendix for completeness).

A more precise location experiment raises the value of information among log-concave location experiments. This is because the precision order strengthens the Lehmann order on such experiments (\citealp*{lehmann_comparing_1988, quah_comparative_2009}), and ordering actions by $\beta(a)$ makes each mean-measurable problem a monotone decision problem. The next result shows making the \emph{prior} more precise reduces the value of location experiments under symmetric distributions.
\begin{proposition}\label{valueinfo}
  Let $f$ and $\tilde f$ be symmetric around $\mu$, and let $g$ be symmetric and log-concave. If $\tilde f$ is more precise than $f$, then $v(\tilde f,g)\le v(f,g)$.
\end{proposition}
The value of an experiment increases in how far the induced posterior means spread from $\mu$. A more precise prior compresses the spread of posterior means on two fronts. The posterior mean reacts less to each signal realization (Corollary~\ref{dualcorollary}), and the prior places less weight on extreme states, so signals are less dispersed.

\section{Conclusion}
We extended the attenuation effect of imprecise information beyond the normal-normal model. Introducing a new precision order based on likelihood-ratio dominance, we showed greater imprecision attenuates the posterior mean toward the prior mean across a broad class of distributions common in economic modeling. The order is not merely sufficient for this attenuation ordering but also necessary, so the precision order is characterized by the effect. The characterization provides a more robust foundation for interpreting attenuated behavior as a consequence of imprecise information, whether observational or cognitive, and pins down symmetry and log-concavity as the essential distributional conditions.

Our results speak to signal-extraction problems more broadly. The median posterior mean inherits the same attenuation ordering under strictly log-concave noise. Applied to common-interest elections under sincere voting, attenuation of the median posterior mean biases majority choice toward the alternative nearer the prior mean, and a more precise signal shrinks the set of states on which the majority errs. Making the prior more precise, rather than the signal, moves the posterior mean in the opposite direction, toward the prior mean, and lowers the value of information.

\bibliography{reference}

\section{Appendix: Proofs}

\begin{lemma}\label{lemapprox}
    A uniform prior density $f$ on $[a,b]$ can be approximated by a sequence of symmetric, log-concave densities $(f_n)_{n\in\mathbb{N}}$ satisfying Assumption~\ref{A1} such that $m(s,f_n,g)\to m(s,f,g)$ for any $s$ and $g$ satisfying Assumption~\ref{A1}. If $f$ is additionally symmetric around 0, all $f_n$ can be taken symmetric around 0. An analogous statement holds for uniform noise densities $g$ on $[a,b]$.
\end{lemma}
\begin{proof}
Define the prior density $f_n$ by
\begin{equation} \label{eqapproxunif}
\log f_n(x)=\begin{cases} 
c, & x\in[a,b]\\
c-n (x-b)^2, & x>b\\ 
c-n (x-a)^2, & x<a, 
\end{cases} \end{equation}
where $c$ normalizes $f_n$ to integrate to $1$. Each $f_n$ is symmetric, log-concave, satisfies Assumption~\ref{A1}, and converges pointwise to the uniform density $f$ on $[a,b]$ as $n\to\infty$. If $f$ is symmetric around 0, each $f_n$ is symmetric around 0. We claim dominated convergence gives
\[m(s,f_n,g)= \frac{\int x f_n(x)g(s-x)dx}{\int f_n(x) g(s-x)dx} \xrightarrow[n\rightarrow\infty]{} \frac{\int x f(x)g(s-x)dx}{\int f(x) g(s-x)dx}=m(s,f,g).\]
Both integrands above are dominated by absolutely integrable functions using $f_n(x)\le1/(b-a)$ and that $g$ satisfies Assumption~\ref{A1}. Finally, the denominator limit is strictly positive, so the ratio converges. The argument for uniform noise on $[a,b]$ is analogous.
\end{proof}

\subsection{Proposition~\ref{propch}}
\begin{proof}
\citet[Proposition 3]{chambers_updating_2012} establish the first statement, that \eqref{eqch} holds when $f$ is symmetric and quasi-concave and $g$ is symmetric around 0 and quasi-concave.

For the second statement, \citet[Proposition 7]{chambers_updating_2012} show that \eqref{eqch} for all uniform $g$ symmetric around 0 implies $f$ is symmetric and quasi-concave. By Lemma~\ref{lemapprox}, each uniform $g$ is a pointwise limit of symmetric, quasi-concave $g_n$ satisfying Assumption~\ref{A1} with $m(s,f,g_n)\to m(s,f,g)$ for all $s$, so \eqref{eqch} extends to uniform $g$.

For the third statement, suppose \eqref{eqch} holds for all symmetric, quasi-concave $f$. Writing $\hat s:=s-\mu_f$ and $\hat f(\cdot):=f(\cdot+\mu_f)$, the hypothesis becomes: for all symmetric around 0, quasi-concave $\hat f$,
\begin{align}\label{eqhat}
\begin{split}
&\forall \hat s\leq 0\colon\quad \hat s\leq m(\hat s,\hat f,g)\leq 0,\\
&\forall \hat s\geq 0\colon\quad \hat s\geq m(\hat s,\hat f,g)\geq 0.
\end{split}
\end{align}
Take $\hat f\sim U[-\sigma,\sigma]$, to which \eqref{eqhat} extends by Lemma~\ref{lemapprox}. Then
\[\forall \sigma>0:\quad 0=m(0,\hat f,g)=\frac{\int_{-\sigma}^{\sigma} x\,g(-x)\,dx}{\int_{-\sigma}^{\sigma} g(-x)\,dx}=\frac{\int_{-\sigma}^{\sigma}(-x)\,g(x)\,dx}{\int_{-\sigma}^{\sigma} g(x)\,dx}.\]
The right-hand side converges to $-\mu_g$ as $\sigma\to\infty$, so $\mu_g=0$. Substituting the identity $m(\hat s,\hat f,g)=\hat s-m(\hat s,g,\hat f)$ (see \eqref{eqduality}) into \eqref{eqhat} yields, for all symmetric around 0, quasi-concave $\hat f$,
\begin{align*}
&\forall \hat s\le \mu_g:\quad \hat s\le m(\hat s,g,\hat f)\le \mu_g,\\
&\forall \hat s\ge \mu_g:\quad \hat s\ge m(\hat s,g,\hat f)\ge \mu_g.
\end{align*}
By the second statement, with $g$ as the prior and $\hat f$ as the noise density, $g$ is symmetric and quasi-concave. Since $\mu_g=0$, $g$ is symmetric around 0.
\end{proof}

\subsection{Theorem \ref{theorem1}}

\begin{proof}
    \textbf{1$\,\boldsymbol{\Rightarrow}\,$2:}
Let $s\ge\mu_f$, the case $s\le\mu_f$ being analogous by symmetry. Recall from \eqref{eqmthm1maintext} the representation $m(s,f,g)=s-\int_0^\infty h(\eps,s)\,\nu_g(\eps)\,d\eps$, where $\nu_g$ is the density on $[0,\infty)$ proportional to $(f(s+\eps)+f(s-\eps))g(\eps)$ and
\begin{equation}\label{eqh}
    h(\eps,s) =\eps\, \frac{f(s-\eps) - f(s+\eps)}{f(s-\eps) + f(s+\eps)}
    = \eps\, \frac{1 - f(s+\eps)/f(s-\eps)}{1 + f(s+\eps)/f(s-\eps)}.
\end{equation}
We claim $h(\eps,s)$ is nonnegative and nondecreasing in $\eps$ on $[0,\infty)$.

\textit{Nonnegativity:} By symmetry of $f$ around $\mu_f$ and quasi-concavity (which is implied by log-concavity and positivity), $f(x)$ is greater the closer $x$ is to $\mu_f$. For $s \ge \mu_f$ and $\eps \ge 0$, the point $s - \eps$ is weakly closer to $\mu_f$ than is $s + \eps$, so $f(s-\eps) \ge f(s+\eps)$. Hence $h(\eps,s) \ge 0$.

\textit{Monotonicity:} By the second expression in \eqref{eqh}, it suffices to show the ratio $f(s+\eps)/f(s-\eps)$ is nonincreasing in $\eps$ on $[0,\infty)$, or equivalently that its logarithm is. By symmetry of $f$ around $\mu_f$, $f(s-\eps)=f(\mu_f +(\mu_f-(s - \eps)))=f(2\mu_f - s + \eps)$, so
\begin{equation*}
    \log\frac{f(s+\eps)}{f(s-\eps)} 
    = \log f(s+\eps) - \log f(2\mu_f - s + \eps).
\end{equation*}
Then
\begin{equation*}
    \frac{d}{d\eps} \log\frac{f(s+\eps)}{f(s-\eps)} 
    = (\log f)'(s+\eps) - (\log f)'(2\mu_f - s + \eps) \le 0,
\end{equation*}
where the inequality uses concavity of $\log f$ together with $s + \eps \ge 2\mu_f - s + \eps$, which holds since $s \ge \mu_f$.

Now compare $g$ with the more precise $\tilde g$. The precision order requires $\tilde g(\eps)/g(\eps)$ to be nonincreasing in $\eps$ on $[0,\infty)$, that is, $g$ likelihood-ratio dominates $\tilde g$ on $[0,\infty)$. Multiplying both densities by the common factor $f(s+\eps) + f(s-\eps)$ preserves likelihood-ratio domination, so $\nu_g$ likelihood-ratio dominates $\nu_{\tilde g}$, and in particular first-order stochastically dominates it. Since $h$ is nondecreasing,
\begin{equation*}
    \int_0^\infty h(\eps,s)\nu_g(\eps)\,d\eps \ge \int_0^\infty h(\eps,s) \nu_{\tilde g}(\eps)\, d\eps,
\end{equation*}
and combining with \eqref{eqmthm1maintext} gives $m(s,f,g) \le m(s,f,\tilde g)$.

\textbf{2$\,\boldsymbol{\Rightarrow}\,$1:}
We first show $g$ and $\tilde g$ are symmetric around 0. Applying \eqref{eqordering} to a log-concave prior $f$ symmetric around 0, so $\mu_f=0$, forces $m(0,f,g)=0$. By Lemma~\ref{lemapprox}, this extends to the uniform prior $f\sim U[-\sigma,\sigma]$. The prior density is constant on $[-\sigma,\sigma]$, so $m(0,f,g)=0$ reduces to ${\int_{-\sigma}^\sigma x\,g(-x)\,dx=0}$ for all $\sigma>0$. Differentiating in $\sigma$, gives $\sigma(g(-\sigma)-g(\sigma))=0$, so $g$ is symmetric around 0. The same argument applies to $\tilde g$.

Next we show $\tilde g$ is more precise than $g$. If not, $\tilde g(x)/g(x)$ has a strictly positive derivative at some $x=s>0$, hence is strictly increasing on some interval $[s-\delta,s+\delta]$ with $\delta>0$ by continuous differentiability. Under the uniform prior $f(x)=\tfrac{1}{2\delta}\mathds 1(x\in[-\delta,\delta])$ and signal $s$, the posteriors are proportional to $g(s-x)$ and $\tilde g(s-x)$ on $[-\delta,\delta]$. Their ratio $\tilde g(s-x)/g(s-x)$ is strictly decreasing in $x$, since $\tilde g/g$ is strictly increasing on $[s-\delta,s+\delta]$. Hence the posterior under $g$ strictly likelihood-ratio dominates that under $\tilde g$, so $m(s,f,\tilde g)<m(s,f,g)$ by \eqref{eqmthm1maintext}, violating \eqref{eqordering} at $s>0=\mu_{f}$. Approximating $f$ by symmetric, log-concave $f_n$ (Lemma~\ref{lemapprox}) with $m(s,f_n,g)\to m(s,f,g)$ and $m(s,f_n,\tilde g)\to m(s,f,\tilde g)$, the strict violation persists for large $n$, so \eqref{eqordering} fails for a symmetric, log-concave prior.
\end{proof}

\subsection{Assumptions on the Prior in Theorem~\ref{theorem1}}
\label{app:prior}

The sufficiency direction of Theorem~\ref{theorem1} assumes the prior $f$ is symmetric and log-concave. This section asks whether either assumption can be relaxed while \eqref{eqordering} continues to hold for all $g$ and $\tilde g$ symmetric around $0$ with $\tilde g$ more precise than $g$. Symmetry of the prior is necessary, and log-concavity cannot be weakened to quasi-concavity. The Online Appendix shows the ordering in \eqref{eqordering} can hold at \textit{some} signal realizations $s$ under neither assumption.

Symmetry of the prior is necessary for \eqref{eqordering}. At $s=\mu_f$, \eqref{eqordering} forces $m(\mu_f,f,g)=\mu_f$. If $m(\mu_f,f,g)=\mu_f$ holds for every $g$ symmetric around 0, hence for every uniform $g$ symmetric around $0$ by Lemma~\ref{lemapprox}, then $f$ is symmetric by \citet[Proposition 6]{chambers_updating_2012}.

The following counterexample shows log-concavity cannot be weakened to quasi-concavity. Consider the prior density (up to normalization)
\[
f(x) \propto
\begin{cases}
2 - |x-1| & \text{if } |x-1| \leq 1, \\
1 & \text{if } 1 \leq |x-1| \leq 10, \\
0 & \text{otherwise.}
\end{cases}
\]
The density is symmetric and quasi-concave around $\mu_f=1$, but the kinks at $|x-1|=1$ prevent log-concavity. At $s=0$, compare $\tilde g$ uniform on $[-1,1]$ with the less precise $g$ uniform on $[-5,5]$:
\begin{equation}\label{eqwrongordering}
    m(s,f,\tilde g) = \tfrac{2}{15} > m(s,f,g) = \tfrac{1}{11}>s=0,
\end{equation}
reversing \eqref{eqordering}. As in Lemma~\ref{lemapprox}, all the densities in this example can be approximated by densities satisfying Assumption~\ref{A1} such that \eqref{eqwrongordering} holds.

The intuition is as follows. The peak of the prior around its mean pulls the posterior mean toward $-1$. Under precise noise, the posterior concentrates near the signal $s=0$ and is sensitive to the local shape of the prior at $0$, so the nearby peak pulls the posterior mean down toward $-1$. Under imprecise noise, the posterior spreads over a wider range, the local shape is averaged out across the broad plateau, and the pull of the peak washes out, leaving the posterior mean closer to the signal.

\subsection{Lemma \ref{lemscale}}\label{applemscal}

\begin{proof}
Since $g$ and $g_\sigma$ are symmetric around $0$, truncating to $[0,\infty)$ gives the densities $2g$ and $2g_\sigma$, and $g_\sigma$ is less precise than $g$ if and only if $2g_\sigma$ is larger than $2g$ in the likelihood-ratio order on $[0,\infty)$. For a nonnegative random variable with density $g$, scaling it by $\sigma$ yields a variable larger in the likelihood-ratio order for all $\sigma>1$ if and only if $\log g(e^x)$ is concave (\citealp*{hu_properties_2004}, Remark 5 (iii)).

For the second claim, let $g$ be log-concave and symmetric around 0. We have
\[\frac{d}{dx}\log g(e^x)=\frac{e^x g'(e^x)}{g(e^x)}.\]
Writing $t=e^x$, the ratio $g'(t)/g(t)=(\log g)'(t)$ is nonincreasing in $t$ by log-concavity of $g$, and nonpositive for $t>0$ because symmetry and quasi-concavity make $g$ nonincreasing on $(0,\infty)$. Hence the product $t\,g'(t)/g(t)$ is nonincreasing in $t>0$ and $\log g(e^x)$ is concave.
\end{proof}

\subsection{Corollary \ref{dualcorollary}}

\begin{proof}
The change of variables $\varepsilon=s-x$ gives \eqref{eqduality}:
\begin{align*}
    m(s,f,g)&=\frac{\int x f(x)g(s-x)dx}{\int f(x) g(s-x)dx}
        =\frac{\int (s-\varepsilon)f(s-\varepsilon)g(\varepsilon)d\varepsilon}{\int f(s-\varepsilon)g(\varepsilon)d\varepsilon}\\
        &=s-\frac{\int \varepsilon\, g(\varepsilon) f(s-\varepsilon)d\varepsilon}{\int g(\varepsilon) f(s-\varepsilon)d\varepsilon}=s-m(s,g,f).
\end{align*}

\textbf{1$\,\boldsymbol{\Rightarrow}\,$2:} Let $f,\tilde f$ be symmetric around $\mu$ with $\tilde f$ more precise than $f$. Then $f_0:=f(\cdot+\mu)$ and $\tilde f_0:=\tilde f(\cdot+\mu)$ are symmetric around $0$ and $\tilde f_0$ is more precise than $f_0$. By Theorem~\ref{theorem1} with $f_0,\tilde f_0$ as noise densities, for every symmetric around $0$ and log-concave $g$,
\begin{align}\label{eqproofprior}
\begin{split}
    \forall s\leq 0\colon \quad &s\leq m(s,g,\tilde f_0)\leq m(s,g,f_0)\\
    \forall s\geq 0\colon \quad &s\geq m(s,g,\tilde f_0)\geq m(s,g,f_0).
\end{split}
\end{align}
By \eqref{eqduality}, \eqref{eqproofprior} is equivalent to 
\begin{align}\label{eqproofprior2}
\begin{split}
    \forall s\leq 0\colon \quad &m(s,f_0,g)\leq m(s,\tilde f_0,g)\leq 0\\
    \forall s\geq 0\colon \quad & m(s,f_0,g)\geq m(s,\tilde f_0,g)\geq 0.
\end{split}
\end{align}
By $m(s,f_0,g)=m(s+\mu,f,g)-\mu$, \eqref{eqproofprior2} is equivalent to \eqref{eqorderingprior}.

\textbf{2$\,\boldsymbol{\Rightarrow}\,$1:} Assume $f$, $\tilde f$ and $\mu$ are such that \eqref{eqorderingprior} holds for every symmetric around 0 and log-concave $g$. With $f_0,\tilde f_0$ as above, reversing the two equivalences of $1\Rightarrow2$ gives \eqref{eqproofprior}. Apply the $2\Rightarrow1$ direction of Theorem~\ref{theorem1} with $g$ as prior and $f_0,\tilde f_0$ as noise densities. The proof of that direction uses only symmetric around $0$, log-concave priors, so $f_0,\tilde f_0$ are symmetric around $0$ with $\tilde f_0$ more precise than $f_0$. Translating back, $f,\tilde f$ are symmetric around $\mu$ with $\tilde f$ more precise than $f$.
\end{proof}

\subsection{Proposition \ref{valueinfo}}

\begin{proof}
Representing a signal structure by its distribution $\tau\in\Delta(\Delta(\R))$ over posteriors $\pi\in\Delta(\R)$, the value equals
\begin{align}\label{eqvaluemeans}
  &\E_\tau\!\Big[\max_{a\in\mathcal A}\E_\pi[u(a,x)]\Big] -\max_{a\in\mathcal A}\int u(a,x)f(x)\,dx\nonumber\\
  ={}&\E_\tau\!\Big[\max_{a\in\mathcal A} \big(\alpha(a)+\beta(a)\,\E_\pi[x]\big)\Big] -\max_{a\in\mathcal A}\big(\alpha(a)+\beta(a)\,\mu\big),
\end{align}
since the action-independent term cancels by Bayes plausibility, $\E_\tau[\E_\pi[\gamma(x)]]=\int\gamma(x)f(x)\,dx$. Thus $v(f,g)$ depends only on the distribution of posterior means.

Without loss, $g$ is symmetric around $0$. If $g$ is symmetric around $y$, set $\hat g(\cdot):=g(\cdot+y)$, which is symmetric around 0. Then $m(s,f,g)=m(s-y,f,\hat g)$ and since a deterministic shift of the signal is invertible by the agent, $v(f,g)=v(f,\hat g)$. Let $H_{f,g}$ be the CDF of the posterior mean under noise $g$ and prior $f$. The signal distribution is the convolution of prior and noise, and the posterior-mean distribution is the pushforward of the signal distribution under $s\mapsto m(s,f,g)$. Define $w\colon\R\to\R$ by \[w(m):=\max_{a\in\mathcal A}\big(\alpha(a)+\beta(a)\,m\big),\] which is convex as a maximum of affine functions. By \eqref{eqvaluemeans}, the value can be written as
\[
  v(f,g)=\int w(m)\,dH_{f,g}(m)-w(\mu).
\]
Bayes plausibility gives $\int m\,dH_{f,g}(m)=\mu$, so adding an affine function to $w$ leaves $v(f,g)$ unchanged. Subtracting a supporting line of $w$ at $\mu$, we may assume $w$ is minimal at $\mu$. Since $f$ and $g$ are symmetric around $\mu$ and $0$, $H_{f,g}$ is symmetric around $\mu$, so $v(f,g)$ is unchanged by replacing $w$ with its symmetrization $\tfrac12(w(m)+w(2\mu-m))$, which is convex, symmetric around $\mu$, and minimal at $\mu$. We may thus assume $w$ has these three properties.

Writing out the distribution $H_{f,g}$,
\[
  v(f,g)=\int\underbrace{\int w\big(m(s,f,g)\big)\,g(s-x)\,ds}_{=:\,W_{f,g}(x)} \,f(x)\,dx- w(\mu).
\]
We decompose the change of the value into two parts:
\[
  v(\tilde f,g)-v(f,g)
   =\underbrace{\int\big(W_{\tilde f,g}(x)-W_{f,g}(x)\big)\,\tilde f(x)\,dx}_{=:A}
   +\underbrace{\int W_{f,g}(x)\,\big(\tilde f(x)-f(x)\big)\,dx}_{=:B}.
\]
First, we prove $A\le 0$. Fix a state $x$. The signal density $g(s-x)$ is unchanged, and by Corollary~\ref{dualcorollary} the posterior mean under the more precise prior is weakly closer to $\mu$ and on the same side of $\mu$: $\mu\le m(s,\tilde f,g)\le m(s,f,g)$ for $s\ge\mu$, symmetrically for $s\le\mu$. Since $w$ is convex and minimal at $\mu$, $w$ is nondecreasing in $|m-\mu|$, so $w(m(s,\tilde f,g))\le w(m(s,f,g))$ for every $s$. So $W_{\tilde f,g}\le W_{f,g}$ pointwise and $A\le 0$.

Second, we prove $B\le 0$. We first show $W_{f,g}$ is nondecreasing on $[\mu,\infty)$ and symmetric around $\mu$. Under log-concave noise the location experiment has the monotone likelihood-ratio property, so $m(s,f,g)$ is nondecreasing in $s$. By symmetry of the prior and noise, $m(\cdot,f,g)$ is antisymmetric around $\mu$, so $m(\mu,f,g)=\mu$ and hence $m(s,f,g)\ge\mu$ for $s\ge\mu$. Therefore $|m(s,f,g)-\mu|$ is nondecreasing in $|s-\mu|$. As $w$ is convex, symmetric around $\mu$, and minimal at $\mu$, $w$ is nondecreasing in $|m-\mu|$, so $w(m(s,f,g))=\psi(|s-\mu|)$ for some nondecreasing $\psi$. With $s=x+\eps$,
\[
  W_{f,g}(x)=\int w(m(s,f,g))\,g(s-x)\,ds
            =\E\big[\psi(|x-\mu+\eps|)\big],\quad \eps\sim g.
\]
Since $\psi$ is nondecreasing, it suffices that $|x-\mu+\eps|$ is first-order stochastically increasing in $x$ on $[\mu,\infty)$. For $0\le c_1\le c_2$ and $t\ge 0$,
\[
  \Pr(|c_2+\eps|>t)-\Pr(|c_1+\eps|>t)
   =\int_{c_1}^{c_2}\big(g(t-c)-g(t+c)\big)\,dc\ge 0,
\]
since $g$ is symmetric and quasi-concave and $|t-c|\le t+c$. Thus, $|c+\eps|$ is increasing in first-order stochastic dominance in $c$. Hence $W_{f,g}$ is nondecreasing on $[\mu,\infty)$, and symmetric around $\mu$ by symmetry of $w(m(\cdot,f,g))$ and $g$.

Both $f$ and $\tilde f$ are symmetric around $\mu$, so each places mass $\tfrac12$ on $[\mu,\infty)$. On $[\mu,\infty)$, $\tilde f$ being more precise means $\tilde f/f$ is nonincreasing, so $\tilde f$ is likelihood-ratio, hence first-order stochastically, dominated by $f$. As $W_{f,g}$ is nondecreasing on $[\mu,\infty)$, \[\int_\mu^\infty W_{f,g}(x)\,\tilde f(x)\,dx\le\int_\mu^\infty W_{f,g}(x)\,f(x)\,dx.\] The inequality holds on $(-\infty,\mu]$ by symmetry of $W_{f,g}$ around $\mu$, so $B\le 0$.

Together, $v(\tilde f,g)-v(f,g)=A+B\le 0$.
\end{proof}

\newpage
\section{Online Appendix}

\subsection{Theorem~\ref{theorem1} without a Symmetric, Log-concave Prior}
\label{apppriorpos}
Both symmetry and log-concavity of the prior enter the proof of Theorem~\ref{theorem1} only through $h(\eps,s)$ in the representation~(3) of the posterior mean. For $s\ge\mu_f$, whenever $h(\cdot,s)$ is nonnegative and nondecreasing on the noise support, the argument of Theorem~\ref{theorem1} yields
\begin{equation}\label{eqorderings}
     s\ge m(s,f,\tilde g)\ge m(s,f,g)
\end{equation}
and symmetrically for $s\le\mu_f$. These properties can hold at some signals under neither symmetry nor log-concavity of the prior, as the following shows.

Suppose the noise density has bounded support $[-\bar\eps,\bar\eps]$, $s\ge\mu_f$, and $f$ is nonincreasing on $[s-\bar\eps,s+\bar\eps]$, the states from which the signal may have originated. For $\eps\ge0$, $f(s-\eps)\ge f(s+\eps)$, and as $\eps$ increases $f(s-\eps)$ rises while $f(s+\eps)$ falls, so $f(s+\eps)/f(s-\eps)$ is nonincreasing in $\eps$. Hence
$$h(\eps,s) = \eps\,\frac{1 - f(s+\eps)/f(s-\eps)}{1 + f(s+\eps)/f(s-\eps)}$$
is nonnegative and nondecreasing on the noise support, and the argument of Theorem~\ref{theorem1} yields~\eqref{eqorderings} at this $s$. A leading case is a quasi-concave prior, nonincreasing to the right of its mode, so the condition holds for any signal far enough above the mode relative to the noise support. If $f$ is in addition symmetric around $\mu_f$, the mode is $\mu_f$ and the condition reduces to $s-\mu_f\ge\bar\eps$, when the signal lies farther from the prior mean than the largest noise realization.

\subsection{Non-log-concave Location-scale Experiments}
The main text gives several examples of commonly encountered symmetric, log-concave distributions. For several symmetric distributions that are not log-concave, $\log g(e^x)$ is nevertheless concave.

For example, the \textit{Student-t} distribution with parameter $\nu>0$ gives
\begin{align*}
    g(x)&\propto \left(1+\frac{x^2}{\nu}\right)^{-(\nu+1)/2}\\
    \Rightarrow \log g(e^x)&=C-\frac{\nu+1}{2}\log(1+\frac{1}{\nu}e^{2x})\\
    \Rightarrow \frac{d^2}{dx^2}\log g(e^x)&=\frac{d}{dx}-\frac{\nu+1}{2}\frac{\frac2\nu e^{2x}}{1+\frac1\nu e^{2x}}=
    -\frac{\nu+1}{2}\frac{\frac4\nu e^{2x}}{(1+\frac1\nu e^{2x})^2}<0,
\end{align*}
and hence has log-concave $g(e^x)$.

A symmetric density with Pareto-type tails, $g(x)\propto(1+|x|)^{-\alpha-1}$ with $\alpha>0$, is the analog of the Laplace distribution with a power-law tail replacing the exponential one. The density is not log-concave, yet
\[\log g(e^x)=C-(\alpha+1)\log(1+e^x),\qquad \frac{d^2}{dx^2}\log g(e^x)=-(\alpha+1)\frac{e^x}{(1+e^x)^2}<0,\]
so $g(e^x)$ is log-concave.

\subsection{Average Posterior Means} \label{secaverage}
Section~\ref{secmedian} characterized comparative statics of the median of the population of posterior means. We now turn to the average. Taking the agents' signals to be independent conditional on the state, the empirical average of their posterior means converges to the average posterior mean conditional on the state,
\[\overline m(x,f,g):=\int m(s,f,g)\, g(s-x)\, ds,\]
as the population grows large. The average is the relevant statistic when only an aggregate is observed, such as total or average consumption or investment.

We first show the average posterior mean lies between the state and the prior mean, extending Proposition~\ref{propch}.
\begin{proposition}\label{propsecondorderexp}
Let $f$ be symmetric around $\mu$ and quasi-concave and $g$ be symmetric around 0 and quasi-concave. Then
\begin{align*}
\begin{split}
&\forall x\leq \mu\colon\quad x\leq \overline m(x,f,g)\leq \mu\\
&\forall x\geq \mu\colon\quad x\geq \overline m(x,f,g)\geq \mu.
\end{split}
\end{align*}
\end{proposition}
The proof pairs each signal with its reflection across $\mu$. By symmetry of the prior and noise densities, $m(\cdot,f,g)$ is antisymmetric around $\mu$, and Proposition~\ref{propch} places each posterior mean between the signal and $\mu$. Quasi-concavity of the noise makes the signal on the same side of $\mu$ as the state more likely than its reflection, which places the average between the state and the prior mean. 

\paragraph{Overconfidence} We now compare agents who misperceive the signal precision. Signals are drawn from the objective signal structure with prior density $f$ and noise density $\hat g$, but an agent who perceives the noise density as $g$ reports the posterior mean $m(s,f,g)$ on observing $s$. Overconfidence in the signal, also called overprecision, is pervasive (\citealp*{moore_trouble_2008}). Generalizing the normal-normal definition of \citet{ortoleva_overconfidence_2015}, we define relative (over)confidence through the precision order: the agent perceiving $g$ is more confident than the agent perceiving $\tilde g$ if $g$ is more precise than $\tilde g$.

When the objective noise density is $\hat g$ and the agent perceives it as $g$, the average posterior mean conditional on state $x$ is
\[\overline m(x,f,g,\hat g):=\int m(s,f,g)\,\hat g(s-x)\,ds.\]
\begin{proposition}\label{propoverconfidence}
Let $f$ be symmetric around $\mu$ and log-concave. Let $\hat g,g,\tilde g$ be symmetric around 0 and quasi-concave. If $g$ is more precise than $\tilde g$, then
\begin{align}
\begin{split}
&\forall x\leq \mu\colon \quad x\leq \overline m(x,f,g,\hat g)\leq \overline m(x,f,\tilde g,\hat g)\leq \mu\\
&\forall x\geq \mu\colon \quad x\geq \overline m(x,f,g,\hat g)\geq \overline m(x,f,\tilde g,\hat g)\geq \mu
\end{split}
\end{align}
\end{proposition}
The average posterior mean of the more confident agent is closer to the true state and farther from the prior mean. The bounds by $x$ and $\mu$ follow as in Proposition~\ref{propsecondorderexp}. For the middle inequality, pairing each signal with its reflection expresses the difference between the two averages as an integral of two nonnegative factors: the gap between the posterior means, signed by Theorem~\ref{theorem1}, and the likelihood advantage of the signal on the state's side, signed by symmetry and quasi-concavity of $\hat g$.

Overprecision trades higher individual variance for lower aggregate bias. A more overprecise agent reacts more strongly to her signal, raising the variance of her posterior mean and making her individual report noisier. The same reactivity moves the average posterior mean closer to the state. Averaging over many agents washes out the added variance and retains the reduced bias, so overprecision is socially valuable when only the accuracy of the average posterior mean matters, as when firms or parties cater to average opinion.

\paragraph{Prior Precision} Greater prior precision moves the average posterior mean opposite to greater signal confidence, toward the prior mean and away from the state.
\begin{proposition}\label{proppriorcomp}
Let $f$ and $\tilde f$ be symmetric around $\mu$ and quasi-concave. Then $\tilde f$ is more precise than $f$ if and only if for every symmetric around $0$ and log-concave $g$,
\begin{align}
\begin{split}
&\forall x\leq \mu\colon\quad x\leq \overline{m}(x,f,g)\leq\overline{m}(x,\tilde f,g)\leq \mu\\
&\forall x\geq \mu\colon\quad x\geq \overline{m}(x,f,g)\geq\overline{m}(x,\tilde f,g)\geq \mu
\end{split}
\end{align}
\end{proposition}
The bounds by $x$ and $\mu$ follow from Proposition~\ref{propsecondorderexp}, applied to $f$ and $\tilde f$ separately. For the middle inequality, pairing each signal with its reflection expresses the difference between the two averages as an integral of two nonnegative factors: the gap between the individual posterior means, signed by Corollary~\ref{dualcorollary}, and the likelihood advantage of the signal on the state's side, signed by symmetry and quasi-concavity of $g$. Necessity of the precision order follows by an argument similar to the necessity direction of Theorem~\ref{theorem1}.

\subsection{Proofs}

\begin{proof}[Proof of Proposition \ref{propsecondorderexp}]
Shifting the prior and the signal by the same amount shifts the average posterior mean by that amount, so it is without loss to assume $\mu_f=0$. Let $x\ge 0$. The case $x\le 0$ is analogous by symmetry of $f$ and $g$.

Splitting the integral defining $\overline m$ at $s=0$,
\begin{align}
\overline m(x,f,g)&=\int_0^\infty\big(m(s,f,g)\,g(s-x)+m(-s,f,g)\,g(-s-x)\big)\,ds\nonumber\\
&=\int_0^\infty m(s,f,g)\big(g(s-x)-g(-s-x)\big)\, ds,\label{eqa}
\end{align}
where we used $m(-s,f,g)=-m(s,f,g)$ due to symmetry of $f$ and $g$. For $s,x\ge 0$, symmetry and quasi-concavity of $g$ give $g(-s-x)=g(s+x)\le g(s-x)$, and Proposition~\ref{propch} gives $0\le m(s,f,g)\le s$. The integrand is therefore nonnegative, proving $\overline m(x,f,g)\ge 0=\mu$.

For the upper bound, symmetry of $g$ around $0$ gives
\begin{equation}\label{eqb}
x=\int_{-\infty}^\infty s\,g(s-x)\,ds=\int_0^\infty s \big(g(s-x)-g(-s-x)\big)\,ds.
\end{equation}
Comparing \eqref{eqa} and \eqref{eqb} term by term, $m(s,f,g)\le s$ together with $g(s-x)-g(-s-x)\ge 0$ shows the integrand of \eqref{eqa} is at most that of \eqref{eqb}. Hence $\overline m(x,f,g)\le x$.
\end{proof}

\begin{proof}[Proof of Proposition \ref{propoverconfidence}]
Without loss assume $\mu=0$. Let $x\ge 0$. The case $x\le 0$ is analogous by symmetry of $f$, $g$, $\tilde g$, and $\hat g$.

The bounds $x\ge\overline m(x,f,g,\hat g)$ and $\overline m(x,f,\tilde g,\hat g)\ge\mu$ follow exactly as in Proposition~\ref{propsecondorderexp}, with the objective noise density $\hat g$ as the weighting density. It remains to show the middle inequality $\overline m(x,f,g,\hat g)\ge\overline m(x,f,\tilde g,\hat g)$. Splitting each defining integral at $s=0$ and substituting $s\mapsto-s$ on the negative part,
\begin{align*}
\overline m(x,f,g,\hat g)&=\int_{0}^{\infty}\big(m(s,f,g)\,\hat g(s-x)+m(-s,f,g)\,\hat g(-s-x)\big)\,ds,\\
\overline m(x,f,\tilde g,\hat g)&=\int_{0}^{\infty}\big(m(s,f,\tilde g)\,\hat g(s-x)+m(-s,f,\tilde g)\,\hat g(-s-x)\big)\,ds.
\end{align*}
Subtracting and using antisymmetry $m(-s,f,g)=-m(s,f,g)$ and $m(-s,f,\tilde g)=-m(s,f,\tilde g)$, which hold by symmetry of the prior and noise densities,
\[
\overline m(x,f,g,\hat g)-\overline m(x,f,\tilde g,\hat g)=\int_{0}^{\infty}\big(\underbrace{m(s,f,g)-m(s,f,\tilde g)}_{\ge 0}\big)\big(\underbrace{\hat g(s-x)-\hat g(-s-x)}_{\ge 0}\big)\,ds\ge 0.
\]
The first factor is nonnegative by Theorem~\ref{theorem1}, since $g$ is more precise than $\tilde g$ and $s\ge\mu=0$. The second factor is nonnegative by symmetry and quasi-concavity of $\hat g$, which give $\hat g(s-x)\ge\hat g(s+x)=\hat g(-s-x)$ for $s,x\ge 0$.
\end{proof}

\begin{proof}[Proof of Proposition~\ref{proppriorcomp}]
\textbf{Sufficiency:} Without loss assume $\mu=0$. We treat $x\ge 0$, the case $x\le 0$ being analogous by symmetry of $f$, $\tilde f$, and $g$.

The bounds $x\ge\overline m(x,f,g)$ and $\overline m(x,\tilde f,g)\ge\mu$ follow from Proposition~\ref{propsecondorderexp}, applied to the priors $f$ and $\tilde f$ respectively (both symmetric around $\mu$ and quasi-concave, with $g$ symmetric and quasi-concave). It remains to show the middle inequality $\overline m(x,f,g)\ge\overline m(x,\tilde f,g)$. Splitting each defining integral at $s=0$ and substituting $s\mapsto-s$ on the negative part,
\begin{align*}
\overline m(x,f,g)&=\int_0^\infty\big(m(s,f,g)\,g(s-x)+m(-s,f,g)\,g(-s-x)\big)\,ds,\\
\overline m(x,\tilde f,g)&=\int_0^\infty\big(m(s,\tilde f,g)\,g(s-x)+m(-s,\tilde f,g)\,g(-s-x)\big)\,ds.
\end{align*}
Subtracting and using antisymmetry $m(-s,f,g)=-m(s,f,g)$ and $m(-s,\tilde f,g)=-m(s,\tilde f,g)$, which hold by symmetry of the priors and the noise density,
\[
\overline m(x,f,g)-\overline m(x,\tilde f,g)=\int_0^\infty\big(\underbrace{m(s,f,g)-m(s,\tilde f,g)}_{\ge 0}\big)\big(\underbrace{g(s-x)-g(-s-x)}_{\ge 0}\big)\,ds\ge 0.
\]
The first factor is nonnegative by Corollary~\ref{dualcorollary}: since $\tilde f$ is more precise than $f$ and $g$ is log-concave, $m(s,f,g)\ge m(s,\tilde f,g)$ for $s\ge\mu=0$. The second factor is nonnegative by symmetry and quasi-concavity of $g$, which give $g(s-x)\ge g(s+x)=g(-s-x)$ for $s,x\ge 0$.

\textbf{Necessity:} If $\tilde f$ is not more precise than $f$, the ordering fails for some symmetric around $0$, log-concave $g$ and some state $x$. Assume $\mu=0$. Then $\tilde f/f$ is not nonincreasing on $(0,\infty)$, so it has a strictly positive derivative at some $x_0>0$, hence is strictly increasing on some interval $[a,b]$ with $0<a<b$ by continuous differentiability (Assumption~\ref{A1}). Set $x:=(a+b)/2$ and take $g$ uniform on $[-\eta,\eta]$ with $\eta:=(b-a)/4$, half the width of $[a,b]$, so that $[x-2\eta,x+2\eta]=[a,b]$.

Under uniform noise the posterior given signal $s$ is the prior restricted to $[s-\eta,s+\eta]$ and normalized, so $m(s,f,g)$ is the mean of $f$ on $[s-\eta,s+\eta]$ and $m(s,\tilde f,g)$ that of $\tilde f$. For every $s\in[x-\eta,x+\eta]$ the window satisfies $[s-\eta,s+\eta]\subseteq[x-2\eta,x+2\eta]=[a,b]$, on which $\tilde f/f$ is strictly increasing. Thus $\tilde f$ restricted to $[s-\eta,s+\eta]$ strictly likelihood-ratio dominates $f$ restricted to the same window, so $m(s,\tilde f,g)>m(s,f,g)$ for every $s\in[x-\eta,x+\eta]$. Averaging against the common weight $g(s-x)$,

\[\overline m(x,\tilde f,g)-\overline m(x,f,g)=\int\big(m(s,\tilde f,g)-m(s,f,g)\big)\,g(s-x)\,ds>0.\]

At $x=(a+b)/2>0=\mu$ the ordering requires $\overline m(x,f,g)\ge\overline m(x,\tilde f,g)$, contradicting the strict inequality above.

Finally, by Lemma~\ref{lemapprox} the uniform $g$ is a pointwise limit of symmetric around $0$, log-concave densities $g_n$ satisfying Assumption~\ref{A1}, with $m(s,\cdot,g_n)\to m(s,\cdot,g)$ for every $s$. The construction gives $g_n(y)=e^{c_n}$ on $[-\eta,\eta]$ and $g_n(y)=e^{c_n}e^{-n(|y|-\eta)^2}$ outside, with $e^{c_n}<1/(2\eta)$. Hence for every $n\ge 1$ and all $y$, $g_n(y)\le \Phi(y):=\tfrac{1}{2\eta}\mathbf 1_{[-\eta,\eta]}(y)+\tfrac{1}{2\eta}e^{-(|y|-\eta)^2}\mathbf 1_{\{|y|>\eta\}}$. Since $|m(s,\cdot,g_n)|\le |s|$ by Proposition~\ref{propch}, the integrand $m(s,\cdot,g_n)\,g_n(s-x)$ is bounded by $|s|\,\Phi(s-x)$, which is integrable because $\Phi$ has Gaussian tails. Dominated convergence gives $\overline m(x,\cdot,g_n)\to\overline m(x,\cdot,g)$ for both $f$ and $\tilde f$, so the strict violation persists for a symmetric around $0$, log-concave $g_n$.
\end{proof}

\end{document}